\documentstyle[preprint,aps,floats,tighten]{revtex}

\begin{document}

\draft

\title{ Macroscopic quantum phases of a deconfined QCD matter at finite density}

\author{ Ji\v{r}\'{\i} Ho\v{s}ek}

\address{  Institute of Nuclear Physics,
\v{R}e\v{z} near Prague, CZ 250 68, Czech Republic}

\date{\today}
\maketitle

\begin{abstract}
Formalism for a unified description of distinct superfluid phases
of a deconfined QCD matter at finite density together with the
phase of spontaneously broken chiral symmetry is presented.
Dispersion laws of the quasiquark excitations in both diamagnetic
and ferromagnetic phases with spontaneously broken chiral symmetry
are exhibited explicitly.
\end{abstract}

\nopagebreak
\vspace*{1.6truecm}

At large enough baryon number densities the hadrons overlap. Due
to the asymptotic freedom of QCD \cite{asfr} in which we trust
such a hadronic system should behave as a weakly interacting
quark-gluon plasma \cite{col}. Restricting ourselves to
temperatures at which the thermal wavelength becomes comparable
with interparticle spacing we can expect behavior of the system
characteristic of a genuine quantum Fermi liquid.

In this paper we consider the chiral $SU(2)$ limit of QCD which is
a good approximation to the real world within a 20\% accuracy.
Numerical value of the quark number density
$n=\frac{2}{\pi^2}p^3_F$ at $T=0$ at which the deconfinement sets
in can be estimated by taking $n=0.72 fm^{-3}$, the quark number
density inside the nucleon. Coherence temperature at which the
quantum behavior sets in is estimated as follows. Average energy
$p$ of a massless quark equals $n_c n_f n_s 3 \frac{1}{2}k_B T$
which defines the thermal wave length
$\lambda_T=\frac{4}{9}\frac{\pi\hbar}{k_B T}$. It should be larger
than $n^{-1/3}$, the average interquark spacing. Hence, below the
line $T=\frac{4}{9}\pi n^{1/3}$ (with $\hbar=k_B$ set equal to one) in the
$(n^{1/3},T)$ QCD phase diagram we expect the behavior of the
system characteristic of a genuine massless Landau Fermi liquid.
For example, its specific heat $C_V\sim T$.

If at the Fermi surface of a Landau Fermi liquid there is an
attraction between two fermions with momenta $\vec{p}_F$
and$-\vec{p}_F$ respectively, the system becomes unstable with
respect to the spontaneous condensation of Cooper pairs. This
happens in the electron Fermi liquids in metals, and turns that
system into a BCS superconductor \cite{BCS}. This happens also in
the Fermi liquid of ${}^3$He, and turns that system into a chain
of distinct superfluid phases \cite{AB}. It is clearly important
to ask : what are the realistic interactions in the QCD Fermi
liquid, whether it is of the Landau type, and whether the Cooper
instability takes place also there.

Unfortunately, there are no experimental data at present, either
real or the lattice ones which would check our considerations.
There are, however, rather solid theoretical arguments that to a
reasonable approximation the effective interactions between
massless quarks in the Landau QCD Fermi liquid are governed by an
$SU(3)_c \times SU(2)_L \times SU(2)_R$ globally invariant local
four-quark Lagrangian ${\cal L}_{int}$. The individual terms in ${\cal L}_{int}$
originate from : (i) massive (due to Debye screening)
chromoelectric gluon exchange; (ii) massive (due to dynamical
Higgs mechanism) chromomagnetic gluon exchange; (iii) quark
interactions with instanton zero modes; (iv) dynamically generated
massive collective excitation exchange. Since at least points (ii)
and (iv) are not under a safe theoretical control it is necessary
to have a method capable of analyzing the phase structure of the
Landau QCD Fermi liquid for a general form of ${\cal L}_{int}$.

Origin of the Cooper-pair instability is understood at present
within the effective field-theory framework both in
nonrelativistic \cite{pol} and relativistic \cite{schaf}
interacting Fermi systems. The local four-fermion couplings listed
above do contain the attractive channels necessary for Cooper
pairing, and the renormalization group determines their running
from the matching point to the infrared. This provides another
good reason for developing a general method of analysis of the
phase structure of a low-temperature system of many fermions
carrying spin, flavor and color. With perturbative forces in mind
the idea of color superfluidity was mentioned already in
\cite{col}, and elaborated in \cite{frau}. Recent papers
\cite{alf1,rapp,berg} deal already with local four-quark
interactions discussed above.

With a general local four-fermion interaction between
quarks carrying spin, isospin and flavor there can be four types
of the superfluid ground-state condensates different from zero due
to Pauli principle (here we ignore the alluring possibility of
spontaneous parity violation within parity-conserving QCD) :
\begin{eqnarray}
v_{(1)} &=& \langle\overline\psi A\tau_{2}\gamma_{5}\psi^{\cal
C}\rangle \, , \label{v1} \\
v_{(2)} &=& \langle\overline\psi
A\tau_{3}\tau_{2}\gamma_{0}\gamma_{3}\psi^{\cal C}\rangle \, ,
\label{v2} \\
v_{(3)} &=& \langle\overline\psi S
\tau_3\tau_2\gamma_5\psi^{\cal C}\rangle \, , \label{v3} \\
v_{(4)} &=& \langle\overline\psi S \tau_2 \gamma_0 \gamma_3
\psi^{\cal C}\rangle \, \label{v4}
\end{eqnarray}
Although we use the convenient Lorentz-covariant notation with
$\psi^{\cal C}=C\overline\psi$ where $C$ is the charge-conjugation
matrix it is clear that the only sacred property of the ground
state here is the translation invariance. We contemplate four
distinct macroscopic quantum phases :
\begin{enumerate}
\item Condensate $v_{(1)}$ with the color-antisymmetric
Clebsch-Gordan  (CG) matrix
$A$ chosen as $A^{ab}=i\epsilon^{ab3}=-(\lambda_{2})^{ab}$
corresponds to a ground-state expectation value of the order
parameter $\Phi^{c}$. It describes a Lorentz scalar, isosinglet
color triplet superfluid.
\item Condensate $v_{(2)}$ corresponds to a ground-state expectation
value of the order parameter $\Phi^{c}_{I;\mu\nu}$. It describes a
color-triplet superfluid which at the same time behaves as
ordinary as well as flavor ferromagnet : $\Phi^{c}_{I;\mu\nu}$ is
an isospin 1 Lorentz antisymmetric tensor field.
\item Condensate $v_{(3)}$ with the color-symmetric CG matrix $S$
chosen as
$S^{ab}=\frac{1}{3}\delta^{ab}-\frac{1}{\sqrt{3}}(\lambda_{8})^{ab}$
corresponds to a ground-state expectation value of the order
parameter $\Phi^{ab}_{I}$. It describes a Lorentz scalar
color-sextet superfluid which at the same time behaves as a flavor
ferromagnet.
\item Condensate $v_{(4)}$ corresponds to a ground-state
expectation value of the order parameter $\Phi^{ab}_{\mu\nu}$. It
describes an isoscalar color-sextet superfluid which at the same
time behaves as an ordinary ferromagnet.
\end{enumerate}
Recent papers \cite{alf1,rapp,berg} are devoted predominantly to
the first case with some estimates \cite{alf1} made also of the
fourth possibility.

It is important to realize that all terms (\ref{v1}-\ref{v4}) are
invariant with respect to global chiral rotations. This implies
that they cannot produce physical, chiral-symmetry-violating quark
masses in the deconfined phase. There is, however, a ground-state
quark-antiquark condensate
\begin{equation}
w=\langle\overline\psi\psi\rangle \label{cond}
\end{equation}
which can do namely this, and it should be considered together
with the quark-quark condensates (\ref{v1}-\ref{v4}). It is well
established that at finite density the local four-quark
interactions considered above can give rise to the quark masses
\cite{bub,alf1} and, by virtue of the Goldstone theorem also to
the massless pions.

In this Letter we present main steps of a generalization of the
field-theory approach developed for superconductivity by Nambu
\cite{nam1}, and for spontaneous breakdown of chiral symmetry in
the nucleon-pion system by Nambu and Jona-Lasinio \cite{nam2}. It
gives almost immediately the dispersion laws of the true fermionic
excitations as exemplified explicitly below. It enables the
(numerical) analysis of the phase structure of the system for a
given set of couplings, as will be clearly seen from the
following. It also provides a systematic way of investigating the
gapless collective Nambu-Goldstone excitations corresponding to
all symmetries spontaneously broken by the condensates
(\ref{v1}-\ref{cond}). This part of the program will be published
separately.

A self-consistent perturbation theory is defined by its bilinear
Lagrangian with all terms (\ref{v1}-\ref{cond}) taken into account :
\begin{eqnarray}
{\cal L}^{'}_{0} & \equiv & {\cal L}_{0}-{\cal L}_{\Sigma}-{\cal
L}_{\Delta}=
 \overline\psi(i\not\!\partial+\mu\gamma_{0})\psi-
  \overline\psi \Sigma \psi -  \nonumber\\
&& \frac{1}{2} \big[ \overline\psi \big(
  A\gamma_{5}\tau_{2}\Delta_{(1)}+
  A\gamma_{0}\gamma_{3}\tau_{3}\tau_{2}\Delta_{(2)} +
  S\gamma_{5}\tau_{3}\tau_{2}\Delta_{(3)}  +
  S\gamma_{0}\gamma_{3}\tau_{2}\Delta_{(4)} \big) \psi^{\cal C} + H.c. \big]
\label{L0'}
\end{eqnarray}
It is assumed that its ground state for $\Delta_{(i)},\Sigma$
different from zero is energetically advantageous with respect to
the naive ground state corresponding to ${\cal
L}_{0}=\overline\psi(i\not\!\partial+\mu\gamma_{0})\psi$. Nonzero
values of $\Delta_{(i)}$ and $\Sigma$ are eventually found by
solving the gap equations expressing the condition that the
lowest-order perturbative contribution of ${\cal
L}^{'}_{int}={\cal L}_{int}+{\cal L}_{\Delta}+{\cal L}_{\Sigma}$,
using the propagator defined by ${\cal L}^{'}_{0}$, vanishes.

Main trick which enables to realize the program using the standard
field-theory methods (first real insight into the problem was
achieved by the BCS-like variational calculation \cite{alf1})
consists of introducing the field
\begin{equation}
q^{a}_{\alpha A}(x)=\frac{1}{\sqrt{2}}
\left(\begin{array}{c}\psi^{a}_{\alpha A}(x)\\ [0.1in]
P^{ab}T_{AB}\delta_{\alpha\beta}\psi^{\cal C}_{b\beta B}
\end{array}\right)
\label{q}
\end{equation}
The matrices $P$ and $T$ defined as $P=e^{i\alpha}A+e^{i\sigma}S,
\ T=(\cos \Theta + i\tau_{3}\sin \Theta)e^{i\phi}\tau_{2}$ have the
property $P^{+}P=T^{+}T=1$. The field $q$ operates in space of
Pauli matrices abbreviated as $\Gamma_{i}$. In terms of $q$ the
Lagrangian (\ref{L0'}) of central interest is
\begin{equation}
{\cal L}_0^{'}=\overline q \left[
\begin{array}{cccc}
\not\!p-\Sigma+\mu\gamma_0 & \; & \; & -\Delta\\[0.1in]
-\gamma_0 \Delta^{+}\gamma_0 & \; & \; & \not\!p-\Sigma-\mu\gamma_0
\end{array}
\right]q\equiv \, \overline q S^{-1}(q)q.
\end{equation}
We have introduced
\[
\Delta \equiv \Delta_{S}\gamma_{5}+\Delta_{V}\gamma_{0}\gamma_{3}
\] where
\[
\Delta_{S}=\Delta_{(1)}e^{-i(\alpha + \phi)}\cos
\Theta-i\Delta_{(3)}e^{-i(\sigma + \phi)}\sin \Theta \]
\[\Delta_{V}=\Delta_{(4)}e^{-i(\sigma + \phi)}\cos
\Theta-i\Delta_{(2)}e^{-i(\alpha + \phi)}\sin \Theta \]
For rewriting any local four-quark interaction in terms of $q$ it
is convenient first to symmetrize all bilinear combinations
$\overline\psi...\psi$ with respect to $\psi$ and $\psi^{\cal C}$,
and then to use $\psi=\frac{1}{\sqrt{2}}(1+\Gamma_{3})q$, and
$\psi^{\cal C}=\frac{1}{\sqrt{2}}(1-\Gamma_{3})P^{+}T^{+}q$.

Calculation of the quasiquark propagator requires some manual
work. Abbreviating
\[
S(p)\equiv \left(
\begin{array}{cc} I & J \\ K & L
\end{array}
\right) \] and imposing the simplifying constraint $Re
\Delta_{S}\Delta^{*}_{V} = 0$ we find
\[
I = \frac{\not\!p_{+}+\Sigma}{D_{+}}[1+\Delta
\frac{\not\!P+M}{D}\gamma_{0}\Delta^{+}\gamma_{0}(\not\!p_{+}+\Sigma)]\]
\[
K =
\frac{\not\!P+M}{D}\gamma_{0}\Delta^{+}\gamma_{0}(\not\!p_{+}+\Sigma)\,
,
\]
where
\begin{eqnarray}
p^{\mu}_{+} &\equiv & ((p_{0}+\mu),\vec{p}\, )\, , \
D_{+}= (p_{0}+\mu)^{2}-\epsilon^{2}_{p}\, ,
\nonumber\\
\epsilon^{2}_{p}&=& \vec{p}^{\, 2}+ \Sigma^{2} \, , \
M= [D_{+}-(|\Delta_{S}|^{2}-|\Delta_{V}|^{2})]\Sigma \, ,
\nonumber\\
P_{3}&=& [D_{+}-(|\Delta_{S}|^{2}-|\Delta_{V}|^{2})]p_{3}\, ,
\nonumber\\
P_{0}&=& D_{+}(p_{0}-\mu)-(|\Delta_{S}|^{2}-|\Delta_{V}|^{2})(p_{0}+\mu)\,
, \nonumber\\
P_{1}&=& [D_{+}-(|\Delta_{S}|^{2}+|\Delta_{V}|^{2})]p_{1}-2|\Delta_{S}||\Delta_{V}|p_{2}\,
, \nonumber\\
P_{2}&=& [D_{+}-(|\Delta_{S}|^{2}+|\Delta_{V}|^{2})]p_{2}+2|\Delta_{S}||\Delta_{V}|p_{1}\,
,   \nonumber\\
D&=& P^{2}-M^{2}=P_{0}^{2}-(\vec{P}^{\, 2}+M^{2}) \, .
\end{eqnarray}
Formulas for $J$ and $L$ follow from those for $K$ and $I$ by
replacements $\pm\mu\leftrightarrow\mp\mu$, and
$\Delta\leftrightarrow \gamma_{0}\Delta^{+}\gamma_{0}$.

For particular case $\Delta_{V}=0$ considered in \cite{alf1}
(without $\Sigma$, i.e., with $\epsilon_{p}=|\vec{p}\, |$)
\begin{equation}
D_{S} = \left[(p_{0}+\mu)^{2}-\epsilon_{p}^{2}\right] \left[ p_0^2- [(\mu
+\epsilon_p)^{2}+|\Delta_S|^2] \right]
\left[ p_0^2- [(\mu-\epsilon_p)^2 +|\Delta_S|^2] \right]
\label{Ds}
\end{equation}
We have verified (the $p_{0}$ integration in the gap equation is
trivially done using Cauchy theorem) that for the instanton-
mediated interaction the gap equation (4.4) of \cite{alf1} is
obtained in an approximation of keeping only the leading term of
the strongest residue. We have also verified that in the vacuum
sector ($\mu=0$) where the formulas greatly simplify our coupled
gap equations for $\Delta_{(1)}$ and $\Sigma$ are in accord with
corresponding formulas of \cite{dia}.

If the interaction is such that the solution $\Delta_S \neq 0,
\Sigma \neq 0$ is energetically favorable, the lowest-order
self-consistent perturbation theory (nonperturbative in the
coupling constants) turns the system of interacting massless
quarks into a system of noninteracting massive quasiquarks with
the dispersion laws given by (\ref{Ds}). The corresponding
specific heat exhibits characteristic exponential behavior. In the
next mandatory step the list of the physical excitations has to be
supplementented with the gapless Nambu-Goldstone collective
excitations interacting in a calculable manner with the massive
quasiquarks.

Analogously, for the particular case $\Delta_{S}=0$ we find the
dispersion laws of the noninteracting quasiquarks in the form
\begin{eqnarray}
D_{V} &=& [(p_{0}+\mu)^{2} - \epsilon_{p}^2] \nonumber\\
&& \left\{ p_{0}^{2}-\epsilon_{p}^{2}-\mu^{2}+ |\Delta_{V}|^{2} -
2 \left[ (\mu^{2}-|\Delta_{V}|^{2})\epsilon_{p}^{2}+
(p_{3}^{2}+\Sigma^{2})|\Delta_{V}|^{2}\right]^{1/2}\right\} \nonumber\\
&&\left\{p_{0}^{2}-\epsilon_{p}^{2}-\mu^{2}+|\Delta_{V}|^{2} +
2\left[ (\mu^{2}-|\Delta_{V}|^{2})\epsilon_{p}^{2} +
(p_{3}^{2}+\Sigma^{2})|\Delta_{V}|^{2}\right]^{1/2}\right\}
\label{Dv}
\end{eqnarray}
It nicely exhibits spontaneous breakdown of the rotational
symmetry in this phase. The formula (\ref{Dv}) enables
straightforward quantitative analysis of the $\Delta_{V}$
formation (together with $\Sigma$), apparently quite involved in
the variational approach. While the $dp_{0}$ integration remains
trivial, the $d^{3}p$ one becomes involved.

Formalism presented above permits a general analysis of the phase
structure of the deconfined QCD matter for any given interaction
between quarks, i.e., not only the local four-fermion one. The
constraint $Re\Delta_{S}\Delta_{V}^{*}=0$ was used merely for an
easy inversion of the general matrix
\[ \not\!P-M+2Re\Delta_{S}\Delta_{V}^{*}[p_{3}\gamma_{0}\gamma_{5}
-(p_{0}+\mu)\gamma_{3}\gamma_{5}+\Sigma\sigma_{12}]
\]
entering the quasiquark propagator $S$.

The model is defined together with a prescription of handling the
three-momentum integrations. With the formfactor
$F(p^{2})=[\Lambda^{2}/(p^{2}+\Lambda^{2})]^{1/2}$ suggested in
\cite{alf1} the dynamically generated quark mass falls down at
large momenta within logarithmic accuracy in accordance with
asymptotic freedom \cite{pag}. For obtaining realistic numerical
estimates of $\Sigma, \Delta$ and $T_{c}$ the correct functional
dependence of these quantities upon the dimensional parameters of
the model may be crucial.

\section*{Acknowledgments}
This work was supported by the grant GACR 2020506. The author is
grateful to INFN for the financial support of his stay in Pisa
where an important part of the work was done, and Adriano Di
Giacomo for his interest and valuable comments.

\end{document}